\documentclass[12pt]{article}
\usepackage{graphicx}
\usepackage{cite}
\newcommand{\be}{\begin{equation}}
\newcommand{\ee}{\end{equation}}
\newcommand{\bea}{\begin{eqnarray}}
\newcommand{\eea}{\end{eqnarray}}
\newcommand{\bean}{\begin{eqnarray*}}
\newcommand{\eean}{\end{eqnarray*}}
\newcommand{\ba}{\begin{array}}
\newcommand{\ea}{\end{array}}

\newcommand{\slashs}[1]{\not{\!#1}}

\newcommand{\no}{\nonumber}
\newcommand{\norsl}{\normalsize\sl}
\newcommand{\norsc}{\normalsize\sc}
\begin{document}
\begin{titlepage}
\title{Lepton Helicity Distributions in\\
            Polarized Drell-Yan Process}
\author{
{\norsc  Jiro KODAIRA${}^{a,\, b}$\, and Hiroshi YOKOYA${}^b$}\\
\\
\norsl  ${}^a$ Deutsches Elektronen-Synchrotron, DESY\\
\norsl  Platanenallee 6, D 15738 Zeuthen, GERMANY\\
\\
\norsl  ${}^b$ Department of Physics, Hiroshima University\\
\norsl  Higashi-Hiroshima 739-8526, JAPAN}
\date{}
\maketitle

\begin{abstract}
{\normalsize

\noindent
The lepton helicity distributions in the polarized Drell-Yan process 
at RHIC energy are investigated.
For the events with relatively low invariant mass of lepton pair
in which the weak interaction is negligible, only
the measurement of lepton helicity can prove the antisymmetric part
of the hadronic tensor.
Therefore it might be interesting to consider
the helicity distributions of leptons to obtain more information
on the structure of nucleon from the polarized Drell-Yan process.
We estimate the QCD corrections at ${\cal O} (\alpha_s)$ level to the
hadronic tensor including both intermediate $\gamma$ and $Z$ bosons.
We present the numerical analyses for different invariant masses
and show that the $u (\bar{u})$ and $d (\bar{d})$ quarks give different
and characteristic contributions to the lepton helicity distributions.
We also estimate the lepton helicity asymmetry
for the various proton's spin configurations.
}
\end{abstract}

\begin{picture}(5,2)(20,-600)
\put(0,-95){DESY 03-009}
\put(0,-110){HUPD - 0301}
\put(0,-125){January 2003}
\put(350,-120){hep-ph/0301228}
\end{picture}
 
\thispagestyle{empty}
\end{titlepage}
\setcounter{page}{1}
\baselineskip 18pt 

\section{INTRODUCTION}

The measurement of the polarized nucleon structure function
$g_1 (x,Q^2)$  by the European Muon Collaboration~\cite{EMC} 
in the late '80s has opened the door to the hadron spin physics.
In the last fifteen years, great progress has been made
both theoretically and experimentally that has considerably
improved our knowledge of the spin structure of nucleon.
Through these developments, hadron spin physics has grown up
as one of the most active fields attracting considerable
attention.
Now our interest has spread out to various processes
to explore the spin structure of hadrons.
In particular, in conjunction with new kind of experiments,
the RHIC spin project, {\it etc.}, we are now in a position
to obtain more information on the structure of hadrons and
the dynamics of QCD.
The spin dependent quantity is, in general, very sensitive to the
structure of interactions among various particles. 
Therefore, we will be able to study the detailed structure of hadrons
based on QCD.
We also hope that we can find some clue to new physics beyond the
standard model through the new experimental data.

It is now expected that the polarized proton-proton collisions
(RHIC-Spin) at BNL relativistic heavy-ion collider RHIC~\cite{RHIC} 
will provide sufficient experimental data to unveil the 
structure of nucleon. 
Therefore it is important and interesting to investigate various 
processes which might be measured in RHIC experiments.
One of those will be the polarized Drell-Yan process~\cite{DY}.

The polarized Drell-Yan process has been studied by many authors
both for longitudinally~\cite{R,W,K1,G1}
and transversely~\cite{RS,JX,VW,CKM,S} polarized case.
The $W,Z$ productions from the polarized hadrons 
are also investigated~\cite{wzp,K2,G2}.
In this article, we discuss the lepton helicity distributions 
from the polarized Drell-Yan process
at the QCD one-loop level.
The lepton helicity distributions carry more information
on the nucleon structure~\cite{ST,CHJR,DS,KSY}
than the \lq\lq inclusive\rq\rq\ Drell-Yan
observable like the invariant mass distribution of lepton pair.
Now let us consider, for simplicity, the 
virtual $\gamma$ mediated Drell-Yan process.
The subprocess cross section $d \hat{\sigma}$ is
written in terms of the hadronic and leptonic tensors as,
\[ d \hat{\sigma} \propto
      \left( W_{\mu\nu}^S + W_{\mu\nu}^A \right)
          \left( L^{S\, \mu\nu} + L^{A\, \mu\nu} \right)
      = W_{\mu\nu}^S L^{S\, \mu\nu}
           + W_{\mu\nu}^A  L^{A\, \mu\nu} \ . \]
The anti-symmetric part of hadronic tensor $W_{\mu\nu}$
contains spin information on the annihilating partons.
However, for observables obtained after summing over the
helicities of lepton or integrating out
the lepton distributions, this anti-symmetric part drops out.
Furthermore, the chiral structure of QED and QCD interactions
tells us that only particular helicity states are selected
for the $q-\bar{q}$ annihilation.
This observation shows that the polarized and unpolarized
Drell-Yan processes are governed by essentially the same
dynamics.
On the other hand, if we measure the lepton helicity
distributions,  we can reveal the whole structure of the
hadronic tensor.
We will show that the $u (\bar{u})$ and $d (\bar{d})$ quarks give
characteristic contributions to the lepton helicity distributions.

The article is organized as follows.
In Sec.2, we reproduce the tree level cross section
to define our conventions.
We present our calculations for the helicity
distributions of lepton at the QCD one-loop level in Sec.3.
We adopt the massive gluon scheme to regularize
the infrared and mass singularities avoiding
the complexity in the treatment of $\gamma_5$.
The scheme dependence will be discussed in Sec.4 
and we will change the scheme
to the $\overline{\rm MS}$ to perform the numerical studies.
We give the numerical results in Sec.5.
Finally, Sec.6 contains the conclusions.
The explicit forms for the subprocess cross sections are
listed in Appendix A.
The invariant mass distribution of lepton pair in the
massive gluon scheme is given in Appendix B which is
used to identify the scheme changing factor.

\section{POLARIZED DRELL-YAN AT TREE LEVEL}

In this section,
we reproduce the tree level result for the helicity distributions of
leptons in the polarized Drell-Yan process to establish our notation.
For the longitudinally polarized Drell-Yan process,
\[ N_A (P_{A},\lambda_{A}) + N_B (P_B,\lambda_{B}) \to
  l (l , \lambda) + \bar{l} ( l' , \lambda') + X \ ,\]
with $\lambda \, (= \pm)$ being the helicity of each particle,
we introduce the parton distributions $f_a^A (x)\,,\, \Delta f_a^A$ by,
\[ f_a^A (x) = f_a^A (x , +) + f_a^A (x , -) \ ,\
    \Delta f_a^A (x) = f_a^A (x , +) - f_a^A (x , -) \ ,\]
where $f_a^A (x , +/- )$ denotes the distribution of parton
type $a$ with positive/negative helicity in nucleon $A$
with positive helicity.
Based on the factorization theorem,
the hadronic cross section for the helicity distribution
of lepton is given as the convolution of the
parton distributions with the hard subprocess
cross section $d\hat{\sigma}^{a b}$ as, 
\bea
  \lefteqn{d\sigma(\lambda_A , \lambda_B ; \lambda) = 
    \sum_{a , b}
     \int_{\tau}^{1} dx_1 \int_{\tau/x_{1}}^{1} dx_2} \no\\
  &\times& \sum_{\lambda_a , \lambda_b}
     \frac{f_a^A (x_1 ) + \lambda_a \lambda_A \Delta f_a^A (x_1 )}{2}
     \ \frac{f_b^B (x_2 ) + \lambda_b \lambda_B \Delta f_b^B (x_2 )}{2}
    \  d\hat{\sigma}^{a b}(\lambda_a ,\lambda_b ; \lambda) 
              \label{generalf}\ ,
\eea
where
\[ \tau = \frac{Q^2}{S} \equiv \frac{(l + l')^2}{(P_A + P_B)^2}\ .\]
The helicity dependent cross section 
$d \hat{\sigma}^{a b} (\lambda_a , \lambda_b ; \lambda)$
for the subprocess,
\[ a (p_a , \lambda_{a}) + b (p_b , \lambda_{b}) \to 
    l (l , \lambda) + \bar{l} (l' , \lambda') + X \ ,\]
is a function of the partonic invariant variables,
\bean
  s &=& (p_a + p_b)^2 = (x_1 P_A + x_2 P_B)^2 = x_1 x_2 S\ ,\\
  t &=& (p_a - l)^2 \ , \quad  u = (p_b - l)^2 \ ,
\eean
and
\[   Q^2 = q^2 = (l + l')^2 \equiv z\,s \ . \]

The tree level polarized cross section for lepton pair
production,
\[  q (p_q , \lambda_q ) + \bar{q} (p_{\bar{q}} , \lambda_{\bar{q}})
       \to l (l , \lambda) + \bar{l} (l' , \lambda') \ , \]
is given by,
\[ d\hat{\sigma}^T =
    \frac{1}{2 N_c^2 s} \left| M^T \right|^2 d \Phi_2 \ , \]
where $d \Phi_2$ is the two particle phase space and
$N_c (=3)$ is the color factor.
The square of the tree amplitude reads,
\be
   \left| M^T \right|^2 =
         \delta_{\lambda_{q}, - \lambda_{\bar{q}}} \,
         \delta_{\lambda , - \lambda'} \,
         N_c \left( \frac{4 \pi \alpha}{Q^2} \right)^2 \,
         \left| f^{\lambda_q \lambda} \right|^2
         L^{\mu\nu}\, W^T_{\mu\nu} \label{Tsquare} \ ,
\ee
where the tree level hadronic and leptonic tensors are defined as,
\be
   W^T_{\mu\nu} = {\rm Tr} (\, \omega_{\lambda_q} \, p_{\bar{q}}\,
             \gamma_{\mu} \, p_q \, \gamma_{\nu}\, ) \ , \quad
    L_{\mu\nu} = {\rm Tr} (\, \omega_{\lambda}\, l \, \gamma_{\mu}
                   \,  l' \, \gamma_{\nu}\, ) \label{TsquareW} \ ,
\ee
and
\[ L^{\mu\nu}\, W^T_{\mu\nu} = 2\, [ t^2 + u^2 + \lambda_q
                \, \lambda ( u^2 - t^2 )] \ . \]
In Eq.(\ref{TsquareW}) and also below,
all momentum $p$ under Tr operator are understood to be $\slashs{p}$
and $\omega_{\lambda} \equiv (1 + \lambda \gamma_5)/2$.
In Eq.(\ref{Tsquare}), $\alpha$ is the QED fine structure constant 
and the quantity $f^{\lambda_q \lambda}$
depends on the fermion couplings to
the photon and $Z$-boson in the following way,
\be
  f^{\lambda_q \lambda} = -\, e_q + Q_q^{\lambda_q}\,
               Q_l^{\lambda}\, \frac{Q^2}{Q^2 - M_Z^2 + i M_Z \Gamma_Z} 
         \label{z-coupling} \ ,
\ee
where $e_q$ is the electric charge of the quark
in units of the electron charge $e$,
$M_Z$ is the $Z$ mass, $\Gamma_Z$ is the $Z$ width.
The helicity labels $\lambda_q$, $\lambda$ of quark and lepton
take $\pm$.
The quark and lepton couplings to the $Z$ boson are
\[ Q_{q,l}^- = \frac{1}{\sin\theta_W \cos\theta_W}
      \left( T^3_{q,l} - \, e_{q,l}\, \sin^2\theta_W \right)\ ,\
   Q_{q,l}^+ = -\, e_{q,l}\,  \frac{\sin\theta_W}{\cos\theta_W} \ ,\]
where $e_l = -1$, $T^3$ is the third component of the isospin and 
$\theta_W$ is the Weinberg angle.
In the center of mass (CM) frame of
annihilating quarks, the cross section becomes,
\be
  \frac{d\hat{\sigma}^T 
     (\lambda_q , \lambda_{\bar{q}} ; \lambda)}{dQ^2 d\cos\theta}
  = \delta_{\lambda_{q}, - \lambda_{\bar{q}}} \,
      \frac{\pi}{2 N_c} \left(\frac{\alpha}{Q^2} \right)^2 \, 
    \left| f^{\lambda_q \lambda} \right|^2       
     \left( 1 + \cos^2\theta + 2 \lambda_q \lambda \cos\theta \right) 
                           \delta(1-z)  \label{xtree} \ ,
\ee
where $\theta$ is the scattering angle of produced lepton.

In Eq.(\ref{xtree}), the third term which depends on the 
helicities of quark and lepton comes from the antisymmetric
part of hadronic tensor $W^T_{\mu\nu}$.
For the observable after taking the spin sum of lepton,
this antisymmetric part appears in the cross section
only through the parity violating $Z$ interaction
in Eq.(\ref{z-coupling}).
Therefore, for the events with \lq\lq small\rq\rq\ values of $Q^2$,
the information from the antisymmetric part will be completely lost.
Furthermore the chiral structure of Standard Model interaction
forces only particular helicity states to participate in the process
as shown in Eq.(\ref{xtree}).
These observations tell us that for the spin summed over final
states and low $Q^2$ events, the polarized and unpolarized
Drell-Yan processes are governed by the essentially the same
dynamics at least for the $q\, \bar{q}$ initiated process.

\section{QCD ONE-LOOP CALCULATION}

The principle result of this section will be the polarized
Drell-Yan cross section at the QCD one-loop level
with the helicity of lepton being fixed.
At the QCD one-loop level, infrared and mass singularities appear
and we regularize them by giving a non-zero mass $\kappa$
to gluon \cite{VW,MGR} to avoid the complexities from 
the treatment of $\gamma_5$
and phase space integrals.
To perform the numerical analyses by using
the known $\overline{\rm MS}$ parameterization for the parton densities, 
we have to change the scheme.
However, it is well known how to do it~\cite{scheme}.
The diagrams to be calculated are given in Fig.1.

The virtual gluon correction (Figs.1a and 1b) to this process
yields,
\be
  \frac{d\hat{\sigma}^V}{dQ^2 d \cos \theta} = 
    \frac{d\hat{\sigma}^T}{dQ^2 d\cos \theta}
  \, \left( \frac{\alpha_s}{\pi} C_F \right) \, 
  \left[ - \frac{1}{2}\, \ln^2 \frac{Q^2}{\kappa^2}
       + \frac{3}{2}\, \ln \frac{Q^2}{\kappa^2}
       - \frac{7}{4} + \frac{\pi^2}{6} \right] \label{vcor} \ ,
\ee
where $\alpha_s = g_s^2 / 4\pi$ is the strong coupling constant
and $C_F = (N_c^2 - 1)/2 N_c$ for ${\rm SU}(N_c)$ of color.

The amplitude for the real gluon emission (Fig.1c and 1d),
\[  q (p_q , \lambda_q ) + \bar{q} (p_{\bar{q}} , \lambda_{\bar{q}})
       \to  l (l , \lambda) + \bar{l}(l' , \lambda') 
       + g(k) \ , \]
is given by,
\bean
  M^R &=& \delta_{\lambda_q , - \lambda_{\bar{q}}} \,
        \delta_{\lambda , - \lambda'} \, 
   \left( - \, \frac{e^2 g_s}{Q^2} \right)\, f^{\lambda_q \lambda}\ 
     \bar{u}_{\lambda} (l) \gamma^{\mu} v_{\lambda'}(l') \\
     & & \times \ \bar{v}_{\lambda_{\bar{q}}} (p_{\bar{q}})
       \left[ \gamma_{\mu} \frac{1}{\slashs{p}_q - \slashs{k}}
         \gamma_{\nu} + \gamma_{\nu}
        \frac{1}{\slashs{p}_q - \slashs{q}} \gamma_{\mu}
        \right]\, T^a \, u_{\lambda_q} (p_q) \, \epsilon^{\nu}_a (k) \ ,
\eean
where $\epsilon^{\nu}_a (k)$ is the polarization vector
of gluon and $T^a$ is the color matrix.
We have defined $q \equiv l + l'$.
\begin{figure}[h]

\vspace{0.3cm}
\begin{center}
\begin{tabular}{cccc}
 & \includegraphics[height=2.4cm,clip]{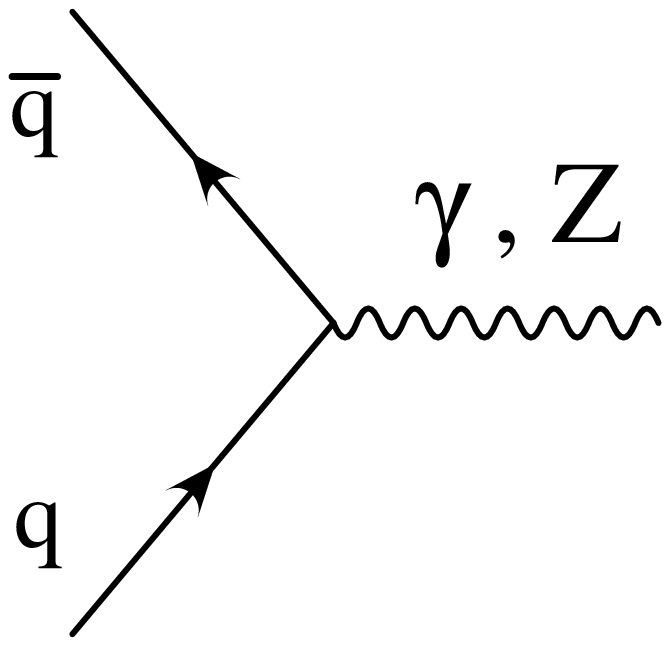} & 
   \includegraphics[height=2.4cm,clip]{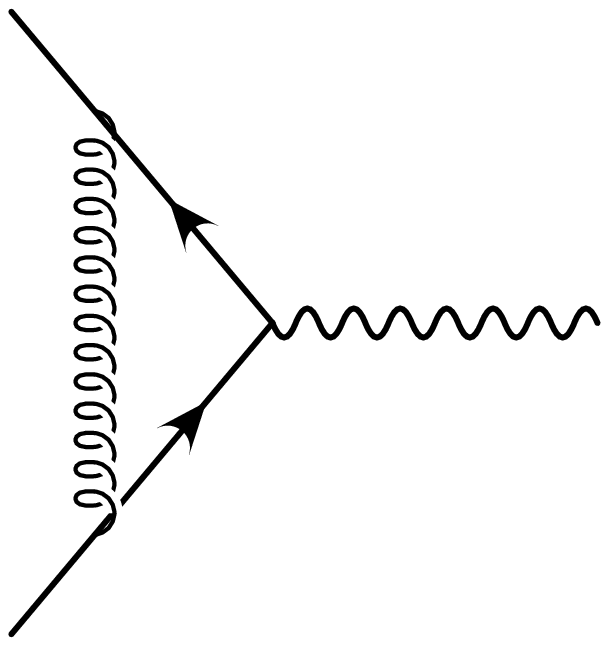} & \\
 & (a) & (b) & \\
 & & & \\
\includegraphics[height=1.6cm,clip]{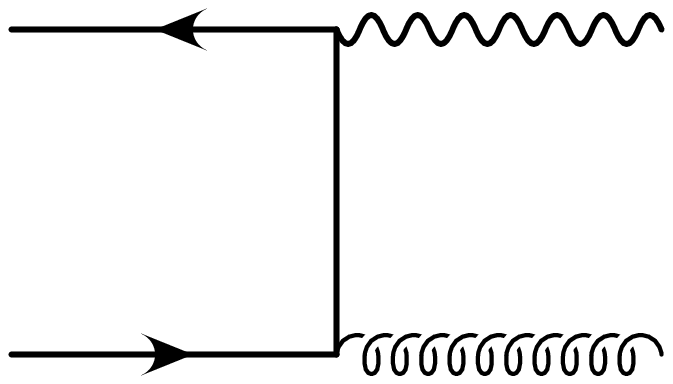} & 
\includegraphics[height=1.6cm,clip]{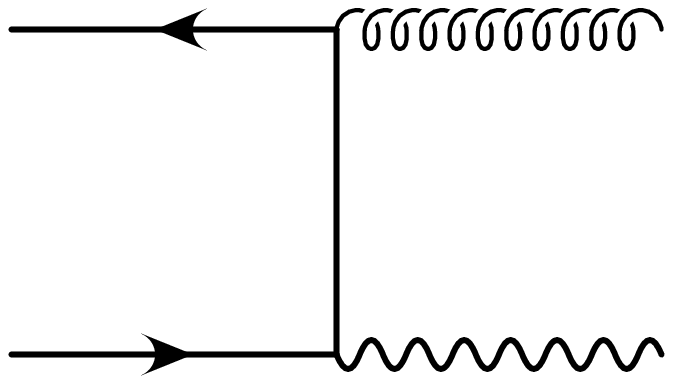} &
\includegraphics[height=1.6cm,clip]{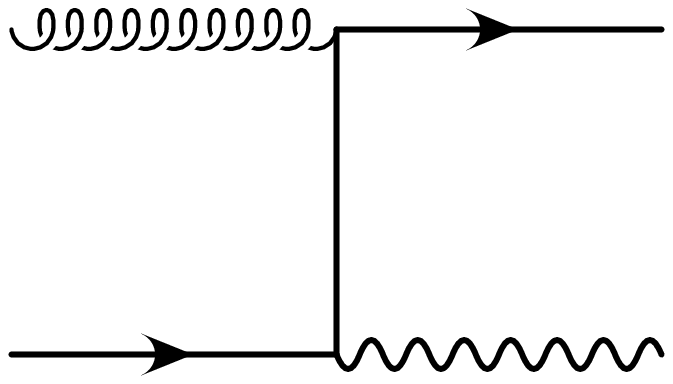} & 
\includegraphics[height=1.7cm,clip]{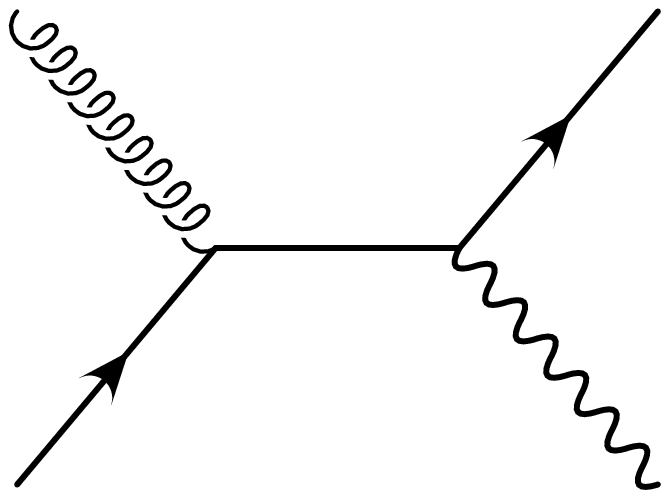}\\
(c) & (d) & (e) & (f) 
\end{tabular}
\caption{Parton-level subprocesses contributing to the Drell-Yan
process at $\mathcal{O} (\alpha_s)$:
(a,b) virtual correction, (c,d) real gluon emission
(e,f) quark-gluon Compton.}
\end{center}
\end{figure}%
The expression for the square of the amplitude given below
has been summed over the spin and colors of unobserved particles.
\[   \left| M^R \right|^2 = \delta_{\lambda_q , - \lambda_{\bar{q}}} \,
       N_c\, C_F\, \left( \frac{4 \pi \alpha g_s}{Q^2} \right)^2 \,
       \left| f^{\lambda_q \lambda} \right|^2\,  
            L^{\mu\nu} \, W^R_{\mu\nu} \ , \]
where
\bea
   W^R_{\mu\nu} &=& \frac{2}{\hat{t}^{2}}
          \left[2 (p_{\bar{q}} \cdot k){\rm Tr}
        (\omega_{\lambda_q} k \gamma_{\mu} p_q \gamma_{\nu})
       - \kappa^2 {\rm Tr}
       (\omega_{\lambda_q} p_{\bar{q}} \gamma_{\mu} p_q \gamma_{\nu})
                              \right]\no\\
    &+& \, \frac{2}{\hat{u}^{2}}
      \left[2 (p_q \cdot k){\rm Tr}
      ( \omega_{\lambda_q} p_{\bar{q}} \gamma_{\mu} k \gamma_{\nu})
            - \kappa^{2}{\rm Tr}
  (\omega_{\lambda_q} p_{\bar{q}} \gamma_{\mu} p_q \gamma_{\nu}) \right]\no\\
  &+& \, \frac{4}{\hat{t}\hat{u}}
   \left[\left( 2 p_q \cdot p_{\bar{q}} - 2 p_q \cdot k
     -2 p_{\bar{q}} \cdot k
                \right) W^T_{\mu\nu}
   + (p_{q \mu} + p_{\bar{q} \mu}) k^{\alpha} W^T_{\alpha\nu}\right.\no\\
  && \left.\hspace{144pt} + \, W^T_{\mu\alpha} k^{\alpha}
         (p_{q \nu} + p_{\bar{q} \nu})
        \right] \label{RsquareW} \ ,
\eea
The tree level hadronic and leptonic tensors
$W^T_{\mu\nu}\,,\,L_{\mu\nu}$ were defined in the previous section
and new invariant variables are introduced,
\[ \hat{t} = (p_q - q)^2 \ ,\, \hat{u} = (p_{\bar{q}} - q)^2 \ .\]
In Eq.(\ref{RsquareW}), we have dropped terms which
do not contribute when $\kappa^2 \to 0$.
It should be noted that the $\kappa^2$ terms in 
the first and second lines
can not be neglected since the phase space integral produces
$1/\kappa^2$ singularity~\cite{MGR}.

The cross section is given by,
\be
  d\hat{\sigma}^R =
    \frac{1}{2 N_c^2 s} \left| M^R \right|^2 d \Phi_3 \label{Xrad}\ ,    
\ee
where $d \Phi_3$ is the three particle phase space.
To integrate out the irrelevant variables in Eq.(\ref{Xrad}),
we take the CM frame of $q \bar{q}$:
\[  p_q = \frac{\sqrt{s}}{2}\, ( 1 , 0 , 0, 1)\ ,\ 
    p_{\bar{q}} = \frac{\sqrt{s}}{2}\, ( 1 , 0 , 0, - 1)\ . \]
Parametrizing the momenta of lepton $l$ and lepton pair $q$ by,
\bean
     l^{\mu} &=& |{\bf l}|\,
     (1 \,,\, \sin\theta\cos\varphi\, , \, \sin\theta\sin\varphi\, ,\,
                \cos\theta) \ ,\\
     q^{\mu} &=& (q^0 \, , \, {\bf q} \, \sin\hat{\theta}\cos\hat{\varphi}\, ,
        \, {\bf q}\, \sin\hat{\theta}\sin\hat{\varphi}\, , 
        \, {\bf q}\, \cos\hat{\theta})\ ,
\eean
where
\bean
   q^0 &=& \frac{s + Q^2 - \kappa^2}{2\sqrt{s}} \qquad , \qquad
                         |{\bf q}|^2 = q^{0 2} - Q^2 \ ,\\
 |{\bf l}| &=& \frac{Q^2}{2 \, \left[ q^0
     - |{\bf q}| \, (\sin\hat{\theta} \sin\theta
        \cos(\hat{\varphi} - \varphi )
      + \cos\hat{\theta}\cos\theta )\right]}\ ,
\eean
it is easy to write the phase space element $d \Phi_3$ as,
\[   d\Phi_3 = \frac{|{\bf q}|\, Q^2 \, dQ^2 \, d\cos\hat{\theta}
           \, d\hat{\varphi} \, d\cos\theta \, d\varphi}
      {32\, (2\pi)^5 \, \sqrt{s}\,
       \left[ q^0 - |{\bf q}|\, (\sin\hat{\theta}\sin\theta
       \cos(\hat{\varphi} - \varphi)
      + \cos\hat{\theta}\cos\theta )\right]^2}\  .\]
After integrating over the angular variables~\cite{VW}
and dropping terms which vanish when $\kappa^2 \to 0$,
we write the inclusive cross section for the polarized lepton
in the following form:

\be
  \frac{d \hat{\sigma}^R (\lambda_q , \lambda_{\bar{q}} ; \lambda)}
                   {d Q^2\, d \cos\theta}
 =  \frac{d \hat{\sigma}^{RT} 
           (\lambda_q , \lambda_{\bar{q}} ; \lambda ; z)}
            {d Q^2\, d \cos\theta}
           \, F^R (Q^2\,,\,z )
       + \frac{d \hat{\sigma}^{RF}
          (\lambda_q , \lambda_{\bar{q}} ; \lambda)}{d Q^2\, d \cos\theta}
           \label{result1}\ ,
\ee
where the first term contains the infrared and mass singularities
as well as terms associated with them and,
\bea
  \lefteqn{\frac{d \hat{\sigma}^{RT} 
          (\lambda_q , \lambda_{\bar{q}} ; \lambda ; z)}
            {d Q^2\, d \cos\theta}  = 
       \delta_{\lambda_{q}, - \lambda_{\bar{q}}} \,     
         \frac{\pi}{2 N_c} \left(\frac{\alpha}{Q^2} \right)^2
    \left| f^{\lambda_q \lambda} \right|^2\cdot \frac{1}{2}} \label{looptree}\\
   &\times&  \left[
        \frac{8 z^2}{(1 + \cos\theta + z (1 - \cos\theta))^4}
      \left\{ (1 + \lambda_q \lambda)(1 + \cos\theta )^2
        + (1 - \lambda_q \lambda) (1 - \cos\theta )^2 z^2 \right\}
                          \right.  \no\\
   &+& \,  \left. \frac{8 z^2}{(1 - \cos\theta + z (1 + \cos\theta))^4}
      \left\{ (1 + \lambda_q \lambda)(1 + \cos\theta)^2 z^2
           + (1 - \lambda_q \lambda) (1 - \cos\theta)^2 \right\} \right] 
                            \no \ .
\eea
with
\bean
  F^R (Q^2\,,\,z) &=&
        \frac{\alpha_s}{\pi} C_F \,
      \left[ \left( \frac{1}{2} \ln^2 \frac{Q^2}{\kappa^2}
            - \frac{\pi^2}{6} \right) \delta(1-z)
      + \frac{1+z^2}{(1-z)}_+ \ln \frac{Q^2}{\kappa^2} \right.\\
   & & \hspace{60pt} 
   \left. + \, 2 (1 + z^2) \left(\frac{\ln{(1-z)}}{1-z}\right)_+
           - 2(1 + z^2) \frac{\ln z}{1-z}
           - (1-z) \right] \ .
\eean
The second term in Eq.(\ref{result1}) 
gives the $\mathcal{O} (\alpha_s)$ finite contribution
whose explicit form is listed in Appendix A.
It should be noted that the first (second) term in Eq.(\ref{looptree})
is just the tree level cross section for the $q\,\bar{q}$ annihilation
with momenta $z p_q$ and $p_{\bar{q}}$
($p_q$ and $z p_{\bar{q}}$) times $z$ which arises from the difference
between the flux normalizations.
These terms express the processes with collinear gluon emissions.
The function $F^R$ can be understood to be a probability
to emit a collinear gluon with momentum $(1-z) p_q$ or
$(1-z) p_{\bar{q}}$ and does not depend on the helicities
of quarks because of the helicity conservation of QCD interaction.

By adding the tree level Eq.(\ref{xtree}) and virtual 
Eq.(\ref{vcor}) contributions to Eq.(\ref{result1}),
the double logarithmic infrared singularities cancel out
and the cross section from the $q\, \bar{q}$ initial states becomes,
\bea
 \frac{d\hat{\sigma}^{T+V+R}}{dQ^2 d\cos\theta}
   &=& \frac{d\hat{\sigma}^{RT}}{dQ^2 d\cos\theta} 
         \ \left[ \left( 1 - \frac{7}{4} \,
       \frac{\alpha_s}{\pi} \, C_F \right) \delta(1-z) 
       +  \frac{\alpha_{s}}{\pi}
         \left( P_{qq}(z) \ln \frac{Q^2}{\kappa^2} 
                 \right.\right. \no\\
   & &  + \, \left. \left.
       C_F \left\{ 2 (1 + z^2) 
         \left( \frac{\ln{(1-z)}}{1-z}\right)_+
        - 2 (1 + z^2) \frac{\ln z}{1-z} - (1 - z) \right\}
                     \right) \right] \no\\
   &+& \frac{d\hat{\sigma}^{RF}}{dQ^2 d\cos\theta} \label{result15} \ ,
\eea
where
\[   P_{qq}(z) = C_F \left( \frac{1+z^2}{(1-z)_+}
                  + \frac{3}{2} \delta(1-z) \right) \ , \]
which is the DGLAP one-loop splitting functions $P_{qq}$. 
To obtain above result, we have used the fact,
\[ \frac{d \hat{\sigma}^{RT} 
           (\lambda_q , \lambda_{\bar{q}} ; \lambda ; z)}
       {d Q^2\, d \cos\theta} \, \delta (1-z)
     = \frac{d \hat{\sigma}^T 
           (\lambda_q , \lambda_{\bar{q}} ; \lambda)}
               {d Q^2\, d \cos\theta} \ .
\]

Finally, the contribution from the quark-gluon Compton
process (Fig.1e and 1f),
\[ q (p_q , \lambda_q) + g (k , h) \to
           l (l , \lambda) + \bar{l}(l' , \lambda') 
       + q (p_q^{\prime} , \lambda_q^{\prime}) \ , \]
where $h$ is the helicity of gluon,
can be calculated in the same way as before.
To avoid singularities in the physical region, the gluon mass
is taken to be $- \kappa^2$~\cite{MGR} and the spin projection
for incoming gluons reads,
\[ \epsilon_{\alpha}(k , h)\, \epsilon_{\beta}^* (k , h)
     = \frac{1}{2}\, \left[ - g_{\alpha\beta}
           + i\, \epsilon_{\alpha\beta\gamma\delta}\, 
             \frac{k^{\gamma}p^{\delta}_q}{k \cdot p_q} \right]\ .\]
The amplitude is,
\bean
  M^C &=& \delta_{\lambda_q , \lambda_q^{\prime}} \,
        \delta_{\lambda , - \lambda'} \, 
   \left( - \, \frac{e^2 g_s}{Q^2} \right)\, f^{\lambda_q \lambda}\ 
     \bar{u}_{\lambda} (l) \gamma^{\mu} v_{\lambda'}(l') \\
     & & \times \ \bar{u}_{\lambda_q^{\prime}} (p_q^{\prime})
       \left[ \gamma_{\mu} \frac{1}{\slashs{p}_q + \slashs{k}}
         \gamma_{\alpha} + \gamma_{\alpha}
        \frac{1}{\slashs{p}_q - \slashs{q}} \gamma_{\mu} \right]\, 
          T^a \, u_{\lambda_q} (p_q) \, \epsilon^{\alpha}_a (k , h) \ ,
\eean
and its square becomes,
\[   \left| M^C \right|^2 = 
       N_c\, C_F\, \left( \frac{4 \pi \alpha g_s}{Q^2} \right)^2 \,
       \left| f^{\lambda_q \lambda} \right|^2\,  
            L^{\mu\nu} \, W^C_{\mu\nu} \ , \]
where
\bean
  \lefteqn{W^C_{\mu\nu} = \frac{1}{s} (1 + \lambda_q h)
         \left(1+\frac{\hat{u}}{\hat{t}}\right){\rm Tr}
 (\omega_{\lambda_q}\, p_q^{\prime}\, \gamma_{\mu} \, k \,  \gamma_{\nu})}\\
  &+&  \left\{ \frac{2 \hat{u}}{s \hat{t}}
           + \frac{1}{s}(1+\lambda_q h)
          + \frac{1}{\hat{t}}(1 - \lambda_q h)\right\}{\rm Tr}
      (\omega_{\lambda_q}\, p_q^{\prime}\, \gamma_{\mu}\, 
                        p_q \, \gamma_{\nu}) \\
  &-&  \frac{1}{\hat{t}} (1 - \lambda_q h)
         \left( 1 + \frac{\hat{u}}{s} \right){\rm Tr}
    (\omega_{\lambda_q}\, k \, \gamma_{\mu}\, p_q \, \gamma_{\nu})
         + \, \frac{4}{\hat{t}} (1 + \lambda_q h)
         p_{q \mu}^{\prime} p_{q \nu}^{\prime}
    + \frac{4}{s} (1 - \lambda_q h) p_{q \mu} p_{q \nu} \\
  &+&  \frac{\kappa^2}{\hat{t}^2} \left(1 - \lambda_q h \,
        \frac{2 Q^2 - s}{s} \right)
         {\rm Tr} (\omega_{\lambda_q}\, (p_q^{\prime} - k )\,
                  \gamma_{\mu}\, p_q \, \gamma_{\nu}) \ ,
\eean
with
\[  s = (p_q + k)^2 \ ,\ \hat{t} = (p_q - q)^2\ ,\ 
       \hat{u} = (k - q)^2 \ . \]
From the cross section formula for this process,
\[ d \hat{\sigma}^C = \frac{1}{2 N_c (N_c^2 - 1) s}
          \left| M^C \right|^2 d \Phi_3 \ , \]
we obtain the polarized lepton distribution and write it
in the same form as Eq.(\ref{result1}),
\be
  \frac{d \hat{\sigma}^C (\lambda_q , h ; \lambda)}
                   {d Q^2\, d \cos\theta}
     =  \frac{d \hat{\sigma}^{CT} (\lambda_q ; \lambda ; z)}
                      {d Q^2\, d \cos\theta}
           \, F^C (Q^2\,,\,z\,,\,\lambda_q \,,\,h)
       + \frac{d \hat{\sigma}^{CF}
          (\lambda_q , h ; \lambda)}{d Q^2\, d \cos\theta}
           \label{result2}\ .
\ee
The first term, in this case, is given by,
\bea
  \lefteqn{\frac{d \hat{\sigma}^{CT} 
          (\lambda_q ; \lambda ; z)}
            {d Q^2\, d \cos\theta}  = 
     \frac{\pi}{2 N_c} \left(\frac{\alpha}{Q^2} \right)^2\,
    \left| f^{\lambda_q \lambda} \right|^2}  \label{comptree}\\
   &\times& \frac{8 z^2}{(1 - \cos\theta + z (1 + \cos\theta))^4}
      \left\{ (1 + \lambda_q \lambda)(1 + \cos\theta)^2 z^2
           + (1 - \lambda_q \lambda) (1 - \cos\theta)^2 \right\} 
                            \no \ ,
\eea
and
\[ F^C (Q^2\,,\,z\,,\,\lambda_q \,,\,h)
   = \left( \frac{\alpha_s}{2 \pi}\right)\, 
  \left[ P^C_{qg} (z; \lambda_q , h)
          \left( \ln \frac{Q^2}{\kappa^2} + \ln\frac{1-z}{z^2}  \right)
     -  \frac{1}{4} (1 + \lambda_q h (1-2z)) \right] \ .\]
The quantity $P^C_{qg}$ is related to the unpolarized and
polarized DGLAP splitting functions $P_{qg}(z)$ and $\Delta P_{qg}(z)$
in the following way,
\[  P^C_{qg} (z; \lambda_q , h) =
       \frac{1}{2} \left( P_{qg}(z) - \lambda_q h \Delta P_{qg}(z)\right)
         = \frac{(1 + \lambda_{q} h)(1 - z)^{2} + 
                           (1 - \lambda_{q} h)z^{2}}{4}\ .\]
Equation (\ref{comptree}) again expresses the tree level
cross section for the $q\,\bar{q}$ annihilation with momenta
$p_q$ and $z k$ where the $\bar{q}$ is emitted collinearly from
the initial gluon.
In contrast to the $q\,\bar{q}$ annihilation case, the function $F^C$
does depend on the helicities of initial quark and gluon
since the emitted antiquark should have the helicity
$\lambda_{\bar{q}} = - \lambda_q$ to annihilate into the electroweak
gauge bosons.
The remaining finite contribution $d \hat{\sigma}^{CF}$
can be found in Appendix A.

The cross section for the $\bar{q}\,g$ subprocess can be obtained
by changing $(\lambda_q , h)$ 
to $(- \lambda_q , - h)$ and $\mu \leftrightarrow \nu$ in $W^C_{\mu\nu}$
in the above formulas.

From these results, the unpolarized cross sections are easily obtained.
As a check, we have reproduced the results of Ref.\cite{MGR}
by neglecting the $Z$ boson contribution.

\section{FACTORIZATION AND SCHEME CHANGE}

Before proceeding to numerical studies, we must factorize
the mass singularities into the parton densities 
in the $\overline{\rm MS}$ scheme
since the two-loop evolution of them which
should be combined with the one-loop corrections
for the hard part, is given in this scheme.

The renormalized parton densities at the factorization scale
$\mu_F^2$ which are relevant to this work
are written as,
\bean
 q (x , \mu_F^2) &=& \int_x^1 \frac{dy}{y}
     \left[ q (y) \left\{ \delta \left( 1- \frac{x}{y} \right)
    + \frac{\alpha_s}{2\pi} \left( P_{qq} \left( \frac{x}{y} \right)
         \ln \frac{\mu_F^2}{\kappa^2} + C_{qq}
   \left( \frac{x}{y} \right) \right) \right\} \right. \\
   && \hspace*{78pt} \left. + \, g (y) \frac{\alpha_s}{2\pi}
        \left( P_{qg} \left( \frac{x}{y} \right) 
         \ln \frac{\mu_F^2}{\kappa^2}
      + C_{qg} \left( \frac{x}{y} \right) \right) \right] \ ,\\
 \Delta q (x , \mu_F^2) &=& \int_x^1 \frac{dy}{y}
      \left[ \Delta q (y) \left\{ \delta \left( 1 - \frac{x}{y} \right)
           + \frac{\alpha_s}{2\pi}
        \left( P_{qq} \left( \frac{x}{y} \right)
                \ln \frac{\mu_F^2}{\kappa^2}
       + C_{qq} \left( \frac{x}{y} \right) \right) \right\} \right. \\
  && \hspace*{48pt} \left. + \, \Delta g (y) \frac{\alpha_s}{2\pi}
       \left( \Delta P_{qg} \left( \frac{x}{y} \right)
        \ln \frac{\mu_F^2}{\kappa^2}
           + \Delta C_{qg} \left( \frac{x}{y} \right) \right) \right] \ ,
\eean
where $C$ and $\Delta C$ are finite functions depending 
on the scheme of factorization.
Since the mass singularity is of universal nature and
not affected by the lepton distributions,
we can find the relation between different regularization schemes
by comparing the results calculated in respective schemes 
for the simpler process.
We have calculated the invariant mass distribution of lepton pair
in our scheme and the results are shown in Appendix B. 
The same quantity has been calculated in the dimensional regularization
scheme in Refs.~\cite{W,G1}.
Note that the result depends on also the prescriptions of $\gamma_5$ 
in this scheme.
We change our massive gluon scheme to the one adopted in Ref.~\cite{G1} 
which can be combined with the known two-loop evolution equations
for the parton densities in Refs.~\cite{V,MV}.
We find,
\bean
 C_{qq}^{\kappa^2 \to \overline{\rm MS}}(z)
       &=& - \, C_F \left( (1 + z^2) \frac{\ln z}{1-z}
             + 2 (1 - z) +\left( \frac{\pi^2}{3}
           - \frac{9}{4} \right) \delta(1-z) \right) \ ,\\
 C_{qg}^{\kappa^2 \to \overline{\rm MS}}(z)
       &=& - \, \left( \frac{(1 - z)^2 + z^2}{2}
               \ln \left( z (1 - z ) \right) + 1 \right) \ ,
\eean
and
\[  \Delta C_{qg}^{\kappa^2 \to \overline{\rm MS}}(z)
       = - \left( \frac{2z - 1}{2} \ln \left( z (1 - z) \right)
               + \frac{1}{2} \right) \ .\]
We must,therefore, factorize these finite terms together with 
the singular terms from the subprocess cross sections.

The final results for the hard part at $\mathcal{O} (\alpha_s)$
level in the $\overline{\rm MS}$ factorization scheme read
for $q \bar{q}$ annihilation from Eq.(\ref{result15}),
\bea
 \frac{d\hat{\sigma}^{q\bar{q}} (\mu_F^2)}{dQ^2 d\cos\theta}
    &=& \frac{d\hat{\sigma}^{RT}}{dQ^2 d\cos\theta}\ 
        \, \left[ \left\{ 1 + 
       \frac{\alpha_s}{\pi} \, C_F \left( \frac{\pi^2}{3} - 4 \right)
              \right\}  \delta(1-z) 
          + \frac{\alpha_s}{\pi} \, 
         \left( P_{qq}(z) \ln \frac{Q^2}{\mu_F^2} \right.\right. \no\\
   &&  + \, \left. \left. 
        C_F \left\{ 2 (1 + z^2) 
         \left( \frac{\ln{(1-z)}}{1-z}\right)_+
    - (1 + z^2) \frac{\ln z}{1-z} + (1 - z) \right\} \right) \right] \no\\
   &+& \, \frac{d\hat{\sigma}^{RF}}{dQ^2 d\cos\theta} \label{cfqq}\ ,
\eea
and for $q g$ Compton process from Eq.(\ref{result2}),
\bea
  \frac{d \hat{\sigma}^{qg} (\mu_F^2)}{d Q^2\, d \cos\theta}
     &=&  \frac{d \hat{\sigma}^{CT}}{d Q^2\, d \cos\theta}\,
     \frac{\alpha_s}{2 \pi}
       \left[  P^C_{qg} (z; \lambda_q , h) \,
           \left\{ \ln \frac{Q^2}{\mu_F^2} 
                + \ln\frac{(1 - z)^2}{z} \right\} \right. \no\\ 
   & &  \hspace*{168pt}
       + \, \left. \frac{1}{4} (1 - 2 \lambda_q h (1-z)) \right] \no \\ 
   &+& \, \frac{d \hat{\sigma}^{CF}}{d Q^2\, d \cos\theta} \label{cfqg}\ .
\eea

\section{NUMERICAL RESULTS}

We are now ready to predict the helicity distributions of lepton
in proton-proton annihilation.
By denoting the scattering angle of lepton in the CM frame of protons
with momenta $P_A$ and $P_B$
by $\theta^{*}$, the relation between
$\cos\theta$ in the CM frame of partons with momenta
$x_1 P_A$ and $x_2 P_B$ and $\cos\theta^*$ reads,
\[ \cos\theta = 
     \frac{x_2 (1 + \cos\theta^* ) - x_1 (1 - \cos\theta^*)}
          {x_2 (1 + \cos\theta^* ) + x_1 (1 - \cos\theta^*)} \ .\]
The hadronic cross section for the helicity distribution of
lepton is finally given by inserting Eqs.(\ref{cfqq},\ref{cfqg})
in terms of $\theta^*$,
into Eq.(\ref{generalf}) and changing the parton densities into
the $\mu_F^2$ dependent ones.
We use the rapidity $y_l$ in stead of the scattering angle
$\cos\theta^*$ of lepton in this section, which is defined by,
\[ y_l = \frac{1}{2}\, \ln\, \frac{1 + \cos\theta^*}{1 - \cos\theta^*}
           \ . \]
We fix the total energy $S$ of the proton-proton system to be
$\sqrt{S} = 500$ GeV which is the planed highest energy at RHIC.
Although our formulas can be applied to the process for arbitrary $Q$,
we have chosen two values $Q = M_Z$ (the $Z$ boson pole) 
and $Q = 10 {\rm GeV}$ as typical examples.  
For the strong coupling constant, we use the standard
two-loop form with $\Lambda_{\rm QCD}$ which is set
to the value used in the parton densities. 
The following values for other parameters of the standard model
are used:
\bean
   \alpha &=& \frac{1}{128}, \quad
    \sin^2 \theta_W = 0.2315\ ,\\
   M_Z &=& 91.187\ {\rm GeV}\ , \quad \Gamma_Z = 2.5\ {\rm GeV} \ .
\eean
Denoting the helicities of initial protons by $P_{A,B}(\pm)$,
there are three cases to be analyzed
corresponding to $P_A (+) P_B (-)$,
$P_A (+) P_B (+)$, $P_A (-) P_B (-)$ configurations.

We have organized this section as follows.
Firstly, we show the numerical results for the helicity
distributions of lepton with respect to $y_l$.
Secondly, we investigate the factorization scale
and the parton parametrization dependence of the cross section.

\subsection{The Lepton Helicity Distribution}

Here we use the $\overline{\rm MS}$ parameterization of
the parton densities in Refs.\cite{PDF1,PDF2}
and take the factorization scale $\mu_F$ and the renormalization
scale $\mu_R$ to be $\mu_F = \mu_R = Q$.
The scale dependence will be discussed later.

We give in Fig.2 the results at $Q = M_Z$
\begin{figure}[h]
\begin{center}
\includegraphics[width=12cm,clip]{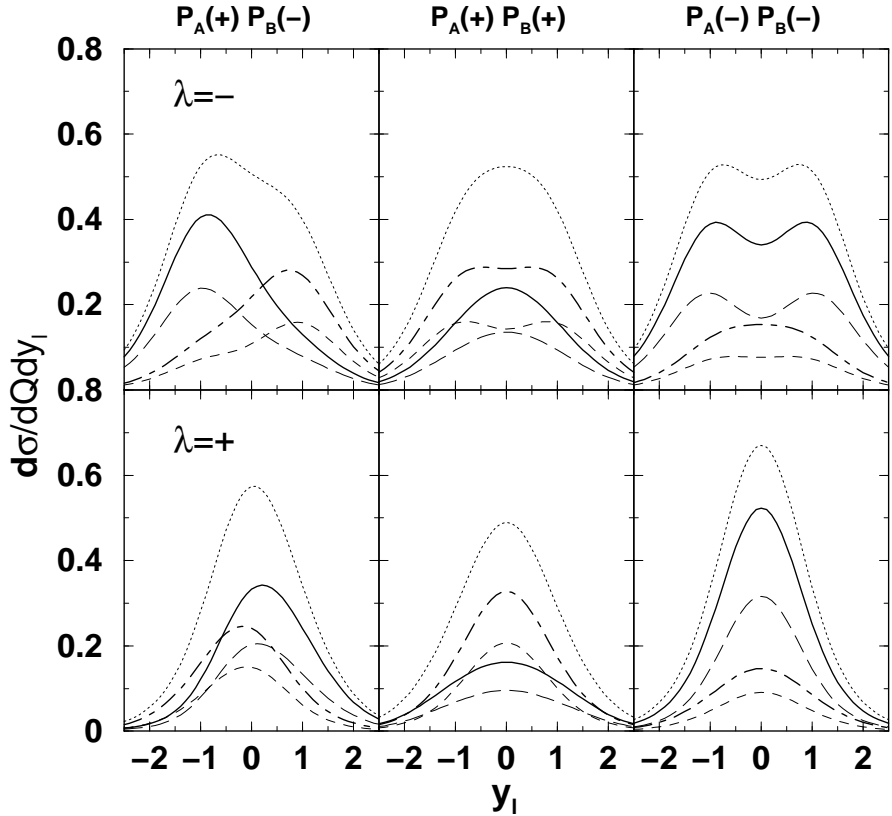}
\caption{The cross sections for leptons with negative and 
positive helicity in the three configurations for the spins of protons
with respect to $y_l$ for $Q = M_Z$.}
\end{center}
\end{figure}%
for both distributions of leptons with negative and positive
helicities in the three $P_A (+) P_B (-)$,
$P_A (+) P_B (+)$, $P_A (-) P_B (-)$ configurations.
The unit of the cross section (vertical axis) is [pb/GeV]. 
The dotted line is the total contribution.
The solid line is the contribution from the $u$-quark,
namely $u \bar{u}$ and $u g (\bar{u} g)$ annihilation sub-processes,
whereas the dot-dashed line from the $d$-quark sector:
we write simply $d$ for the $d$ and $s$ quarks.
The long dashed (dashed) line is the contribution
from the tree level $u \bar{u}$ ($d \bar{d}$).
We plot in Fig.3 the similar results for $Q = 10$ GeV.
Some comments are in order for these results.
Firstly, the main effect of QCD correction is just an enhancement
of the tree level cross section: the $K$ factor is
$1.5 \sim 1.8$.
It does not change significantly the shape of the lepton
distributions from the tree level one.
\begin{figure}[h]
\begin{center}
\includegraphics[width=12cm,clip]{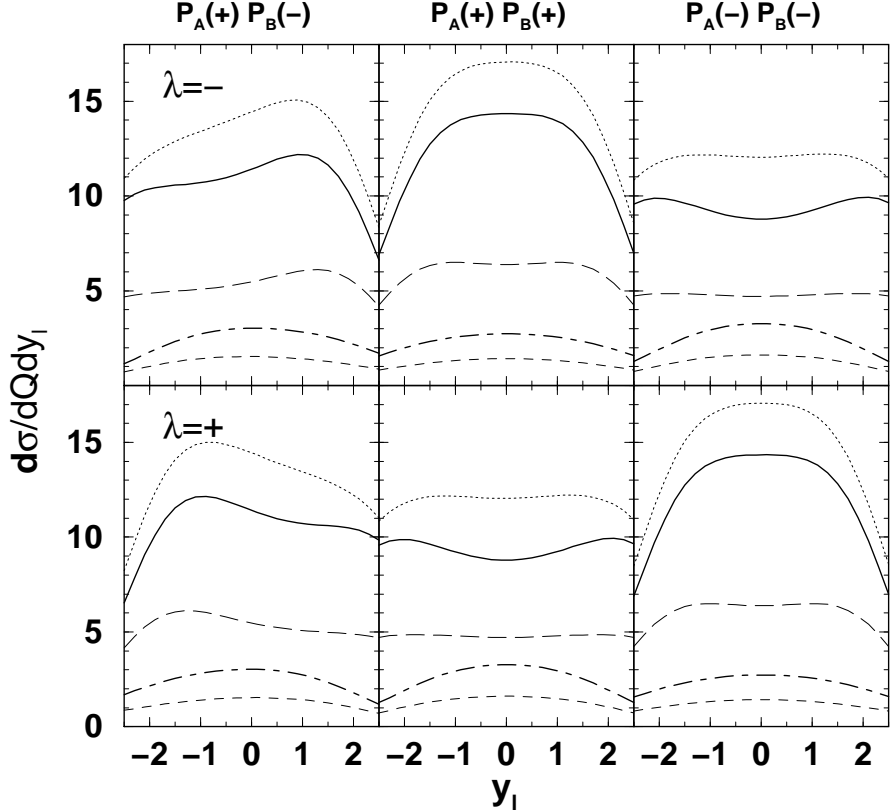}
\caption{The cross sections for leptons with negative and 
positive helicity in the three configurations for the spins of protons
with respect to $y_l$ for $Q = 10$ GeV.}
\end{center}
\end{figure}%
Secondly, we see that the $u$ and $d$ quarks give different
and characteristic contributions to the helicity
distributions for lepton.
In particular, the case of the lepton distribution
in the $P_A(+) P_B(-)$ configuration for $Q = M_Z$ deserves some
explanations.
Our numerical results show that in the negative rapidity
region, the leptons with negative helicities
are mainly produced and those come from the $u$-quark annihilation.
On the other hand, in the positive rapidity region, leptons with both
helicities are produced, however
the leptons with negative helicities come mainly from 
the $d$-quark annihilation subprocesses.
These features can be understood
intuitively by observing the following aspects.
(1) The polarized quark distributions in the polarized proton
$P(+)$ roughly satisfy for the relevant scale $Q^2 = M_Z^2$
and the momentum fractions of partons $x_1 x_2 = M_Z^2 / S$,
\[ u (\uparrow) \gg d (\downarrow) \sim u (\downarrow)
           \gg d (\uparrow) \gg \bar{u} (\uparrow , \downarrow)
         \sim  \bar{d} (\uparrow , \downarrow) \ ,\]
where $\uparrow$ and $\downarrow$ denote the quark's spin parallel
and anti-parallel to the parent proton's spin.
This relation implies the dominant subprocesses in the $P_A (+) P_B (-)$
configuration to be 
(i) $u_A (\uparrow) \bar{u}_B$, (ii) $\bar{u}_A u_B (\uparrow)$,
(iii) $d_A (\downarrow) \bar{d}_B$ and (iv) $\bar{d}_A d_B (\downarrow)$.
(2) From the angular momentum conservation,
the spin of produced $Z$ boson is aligned to $P_A$ ($P_B$) direction
for $u_A \bar{u}_B$ and $\bar{u}_A u_B$ 
($d_A \bar{d}_B$ and $\bar{d}_A d_B$) annihilations and
the negative (positive)
helicity lepton from the $Z$ decay has higher probability to be produced
in the opposite (same) direction of $Z$ boson's spin.  
(3) The third point to be noted is that the $V-A$ coupling is larger
than the $V+A$ coupling for the quark-$Z$ boson interaction.
This suggests that among four subprocesses in (1),
(ii) and (iii) eventually dominate the process.
(4) Finally, since the momentum fractions of quarks are bigger than those of
anti-quarks, the distributions of negative helicity lepton
from (ii) and (iii) are Lorentz boosted to the negative 
and positive rapidity regions respectively.
The fact above (3) also explains that the $d$ ($u$) quark contribution is
larger than that of $u$ ($d$) quark
for the $P_A(+) P_B(+)$ ($P_A(-) P_B(-)$) configuration.
For these configurations, the lepton distribution is 
symmetric around $y_l = 0$ as it should be.

At the lower value of $Q$ ($Q = 10$ GeV),
the $Z$ boson contribution can be safely neglected
and the electromagnetic interaction becomes dominant.
In this case, the situation is not so simple as the previous one
($Q = M_Z$).
At first sight, only the point (3) above should be dropped.
If so, the subprocess (i) and (ii) had dominated the process
which leaded to the similar distributions as those for the $Q = M_Z$ case.
However, the distribution takes somehow different shape due to the
following reasons.
For $Q = 10$ GeV, the partons with very small momentum
fractions participate in the process ($x_1 x_2 = 10^2 / S$).
Therefore, for the \lq\lq dominant\lq\lq\  processes 
for the $Q =M_Z$ case, (i) and (ii),
the Lorentz boost effect is fairly large and the Jacobian
factor $J = \sin^2\theta^*$ apparently diminishes the
cross section in the $y_l$ distributions.
Furthermore at this scale, all subprocesses seem to
contribute equally to the cross section and the partons with small $x$
have less information on the spin of the parent proton. 
One can see, nonetheless, the contribution from the
antisymmetric part of the hadronic tensor
which causes the asymmetric distributions of lepton for
the $P_A(+) P_B(-)$ configuration.  

We also estimate the lepton helicity asymmetry $A_l$ 
which is defined by,
\[ A_l \equiv \frac{d\sigma(\lambda = -1) - d\sigma (\lambda = +1)}
    {d\sigma(\lambda = -1) + d\sigma(\lambda = +1)} \ .\]
This asymmetry is plotted in Fig. 4 for the 
three spin configurations for protons.
The upper three graphs show the asymmetry for the $Q = M_Z$ case
and the lower three for the $Q = 10$ GeV case.
\begin{figure}[h]
\begin{center}
\includegraphics[width=12cm,clip]{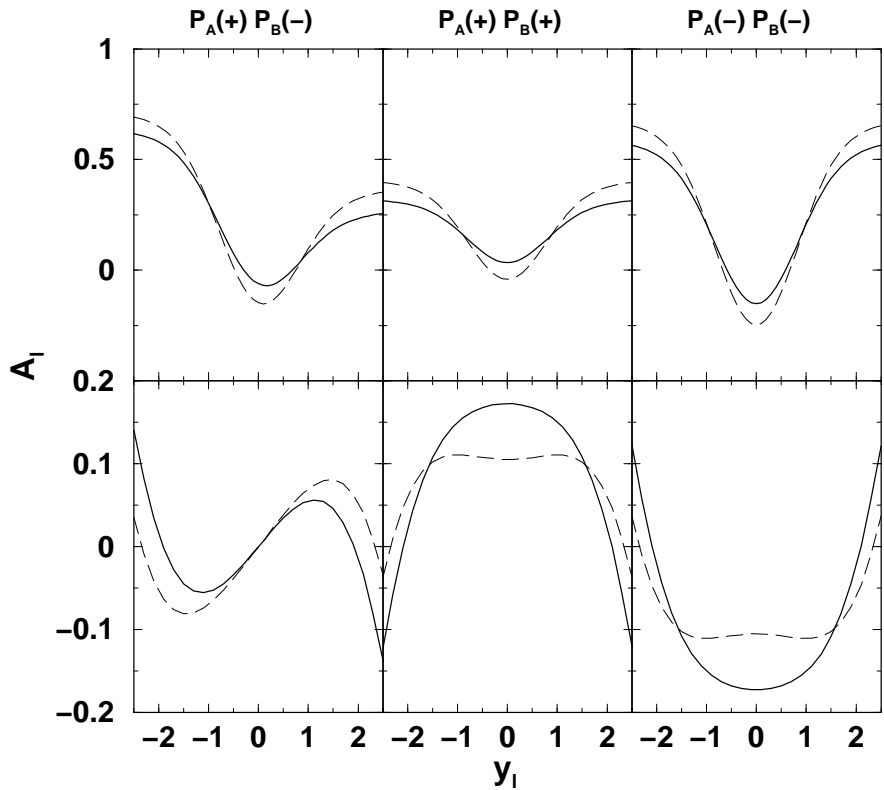}
\caption{The helicity asymmetry
in the three configurations for the spins of protons
with respect to $y_l$.}
\end{center}
\end{figure}%
The solid line is the $\mathcal{O} (\alpha_s)$ result and 
the long-dashed line is the tree level one.
In these results, we recognize that there are non negligible effects from
the QCD higher order corrections other than just the
enhancement effects of tree level results.
This asymmetry amounts to around 20 - 30\% in the 
central rapidity region for $Q = M_Z$.
The asymmetry for $Q = 10$ GeV is much smaller than
that for $Q = M_Z$.
This is consistent with the previous observation that
partons with small $x$ which do not have enough spin information
of parent proton, contribute to the process at smaller $Q$.

\subsection{The Scale and Parametrization Dependence}

The physical quantities such as the cross sections or
asymmetries are obviously independent of the
renormalization and the factorization scales.
However the truncation of the perturbative series
at a fixed order induces the dependence on these
scales which leads to uncertainty in the theoretical
predictions.
We study the scale dependence of the
cross section and asymmetry 
by comparing the results obtained with three different
choices for the scale: 
$\mu_F^2 = 2 \, Q^2$, 
$\mu_F^2 = Q^2$ and $\mu_F^2 = 0.5 \, Q^2$.
We will also mention the dependence on the various
parton parametrizations in this subsection.

We give the results only for the case of $P_A(+) P_B(-)$ configuration
with $Q = M_Z$ since the results 
for other spin configurations of protons and the value of $Q$,
are found to be essentially the same.
\begin{figure}[h]
\begin{center}
\includegraphics[width=11cm,clip]{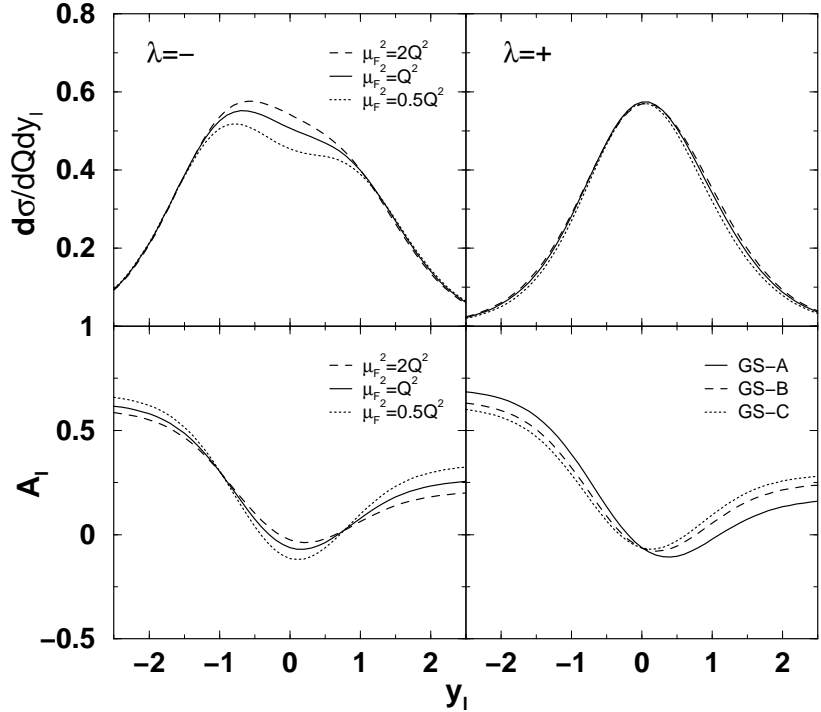}
\caption{The scale and the parton parametrization
dependence.}
\end{center}
\end{figure}%
The scale dependence of the lepton distributions 
(the upper two graphs) as well as the asymmetry (the lower left graph)
is shown in Fig. 5 where the dashed line corresponds
to $\mu_F^2 = 2 \, Q^2$, the solid line to $\mu_F^2 = Q^2$
and the dotted line to $\mu_F^2 = 0.5 \, Q^2$.
In this calculation, we use the parton densities in Ref.~\cite{PDF1}. 
One can see that the variation is less than 10\% in the
central rapidity region.

For the parton parametrization dependence, we calculate
the asymmetry with the three sets in Ref.~\cite{PDF3}
since the difference appears more clearly in the asymmetry
than in the cross sections.
The results are shown in Fig. 5 (the lower right graph).
The solid line, dashed line and dotted line
correspond to GS-A, GS-B and GS-C respectively.
The dependence on the parton parametrizations
is similar to the scale dependence in size and
not so large.
The reason will be that the process is dominated by the
quark and anti-quark annihilation in the RHIC energy region
and the gluon initiated Compton subprocess,
in which the ambiguity of gluon distribution will appear, 
gives a tiny correction to the cross section.

\section{CONCLUSION}

We have presented a complete calculation 
at the  ${\cal O} (\alpha_s)$ order in QCD of
the lepton helicity distributions
in the polarized Drell-Yan process. 
We have numerically analyzed the cross section at
$Q = 10$ GeV and $Q = M_Z$
and pointed out that the $u (\bar{u})$ and $d (\bar{d})$ quarks 
give different and characteristic contributions to the lepton 
helicity distributions which deserve some theoretical
interests.
The QCD corrections mainly enhance the tree level
cross sections and this fact can explain qualitatively
the lepton helicity distributions from the various
proton's spin configurations. 
We have defined and estimated the lepton helicity
asymmetry which amounts to around 20-30\%.

We have also studied the dependence of the cross section
and the asymmetry on the scale and parton
parametrization.
Since the $q\,\bar{q}$ subprocess is dominant in the RHIC
energy region, we will not be able, unfortunately, to
find clear difference between various parton parameterizations
which have big ambiguities in the gluon distributions.

From the experimental point of view,
it seems rather difficult to measure the helicity of
produced muon and/or electron in the Drell-Yan process.
However, if we can observe the $\tau$ lepton produced from
the Drell-Yan process and its decay,
we will be able to compare the experimental data and theoretical
prediction presented in this paper.
We believe that our calculations and numerical studies
may provide a complementary information to the
existing results for the invariant mass and rapidity or $x_F$ 
distributions of lepton pair in the polarized proton-proton
collisions.

\newpage
\begin{center}
ACKNOWLEDGMENTS
\end{center}

We are grateful to Werner Vogelsang for useful discussions 
and suggestions.
We would like to thank T. D. Lee and W. Vogelsang for hospitality
and support extended to them at RIKEN-BNL Research Center
where part of this work was performed.  
J. K. thank J. Bl\"umlein for hospitality
and support extended to him at DESY where this work has 
been completed. 
The work of J. K. was supported
in part by the Monbu-kagaku-sho Grant-in-Aid
for Scientific Research No. C-13640289.

\vspace{1cm}
\begin{center}
APPENDIX A: THE CROSS SECTION FORMULAE
\end{center}

The explicit forms for the finite contributions to the
cross sections are listed in this appendix.
We use the abbreviation $c \equiv \cos\theta$.

\vspace{6pt}
\noindent
1. The real gluon emission process.
\bean
  \lefteqn{\frac{d\hat{\sigma}^{RF}
                        (\lambda_q , \lambda_{\bar{q}} ; \lambda)}
    {dQ^2 d\cos\theta} = 
      \delta_{\lambda_q , - \lambda_{\bar{q}}} \,
       \frac{\pi}{2 N_c} \left( \frac{\alpha}{Q^2} \right)^2
       \left| f^{\lambda_q \lambda} \right|^2 \,     
         \left(\frac{\alpha_s}{\pi} C_F \right) }\\
   &\times& \biggl[ 4z^{2} \frac{1+z^{2}}{1-z} 
         \biggl\{ \frac{(1+\lambda_q \lambda ) (1+c)^2 z^2
         + (1 - \lambda_q \lambda) (1-c)^2}
         {(1-c+z(1+c))^{4}}
           \ln{ \left( \frac{(1 - c + z (1+c))^2}{4z}\right)} \\
   && \hspace*{48pt} + \, \frac{(1 + \lambda_q \lambda) (1 + c)^2
          + (1 - \lambda_q \lambda) z^2 (1 - c)^2}
              {(1 + c + z(1-c))^4}
        \ln {\left(\frac{(1 + c + z(1-c))^2}{4z}\right)} \biggr\} \\
   && - \, 2z(1-z) + \frac{4z(1-z)(1+z^2)
             \left\{ ( 1 + z)^2 - c^2 (1 + 6z + z^2) \right\}}
            {(1+c+z(1-c))^2 (1 - c + z(1+c))^2} \\
   && + \, \frac{8 \lambda_q \lambda z^2 (1-z)c}
             {(1+c+z(1-c))^3 (1-c+z(1+c))^3}
            \left\{ (1+z)^2 (7-4z+7z^2) \right.\\
   && \hspace*{48pt} \left.- \, 2 c^2 (3+14z+10z^2+14z^3+3z^4)
           - c^4 (1-z)^2 (1+z^2) \right\} \\
   && + \, \frac{2z(1-z)}{3(1+c+z(1-c))^4 (1-c+z(1+c))^4} \\
   && \hspace*{24pt}
        \times \, \left\{ -(1+z)^4 (1+14z-18z^2+14z^3+z^4) \right.\\
   && \hspace*{48pt} + \, 2c^2(1+z)^2 (1 + 4z + 239 z^2 - 56 z^3
        + 239z^4 + 4z^5 + z^6) \\
   && \hspace*{48pt} + \, 8 c^4 z (3-45z-151z^2
        - 118z^3 - 151z^4 - 45z^5 + 3z^6) \\
   && \hspace*{48pt} - \, 2 c^6 (1-z)^2 (1 + 8z + 71z^2
                  +48z^3 + 71z^4 + 8z^5 + z^6) \\
   && \hspace*{168pt} \left. + \, c^8 (1-z)^4 (1-2z+6z^2-2z^3+z^4)
                     \right\} \biggr] \ .
\eean

\noindent
2. The quark gluon Compton process.
\bean
  \lefteqn{\frac{d\hat{\sigma}^{CF}
          (\lambda_q , h ; \lambda)}{dQ^2 d\cos\theta}
     = \frac{\pi}{2 N_c} \left( \frac{\alpha}{Q^2} \right)^2
       \left| f^{\lambda_q \lambda} \right|^2 \,     
         \left(\frac{\alpha_s}{\pi} \right) }\\
  &\times& \left[ 4\, z^{2} \, P^C_{qg}(z; \lambda_{q},h)
          \, \frac{(1+\lambda_{q}\lambda)(1 + c)^2 z^2
          + ( 1-\lambda_{q}\lambda )(1 - c)^2}{(1-c+z(1+c))^{4}} \right.\\
  && \hspace*{216pt} \times \, \ln \left(\frac{(1-c+z(1+c))^2}{4z} \right) \\ 
  && \left. + \, \frac{2 z (1 - z)(1 - z^2 - c (1 + 6 z + z^2))}
         {(1 + c + z(1-c))(1 - c + z(1 + c))^3} \right. \\
  && \hspace*{24pt} \left. \times \, 
           \biggl\{ \left((1 - \lambda_q \lambda)(1 - c)
         + (1 + \lambda_q \lambda) z (1 + c) \right)
             P^C_{qg} (z ; \lambda_q , h) \right. \\
  && \hspace*{72pt} \left. -\, \frac{z}{2} \left( (1 - \lambda_q h) z (1-c)
      + (1 + \lambda_q h)(1 + c + z(1 - c)) \right) \biggr\} \right. \\
  && \left. + \, \frac{1}{4}(1 + \lambda_q \lambda)(1 - \lambda_q h) z^2
           \left( 1 - z - c \ln z \right) \right. \\
  && \left. + \, \frac{2\, z\, (1 - z^2)\,  c \, P^C_{qg}(z; \lambda_q , h)}
              {(1 + c + z(1-c))(1 - c + z(1 + c))}
           + G_S + \lambda_q \lambda G_A \right] \ ,
\eean
where,
\bean
 G_S &=& \frac{z(1 - z)}{12 (1 + c + z(1-c))(1 - c + z(1 + c))^4} \\
  &\times& \left[
        (1 - z^2) \left( (1 - \lambda_q h)
          (1 + 5z + 21 z^2 - 7 z^3 - 2 z^4) \right.\right. \\
  && \left.\hspace*{144pt} - \, 2 (1 + \lambda_q h)
                    (1 + 2z - 42 z^2 + 2 z^3 + z^4) \right) \\
  && - \, c \left( (1 - \lambda_q h)
         (4 + 16z + 61z^2 + 98z^3 - 34z^4 + 2z^5 - 3z^6) \right. \\
  && \left.\hspace*{144pt} - \, 4 (1 + \lambda_q h)
              (1-3z^2-92z^3-3z^4+z^6)\right) \no\\
  && + \, c^{2} \left( (1 - \lambda_q h)
        (5 + 17z + 65z^2 + 208z^3 + 13z^4 - 17z^5 - 3z^6) \right. \\
  && \left.\hspace*{144pt} - \, 72 (1 + \lambda_q h) z^2 (1-z^2) \right) \\
  && - \, 2c^3 \left( (1 - \lambda_q h) z
        (4 + 13z + 38z^2 + 16z^3 - 2z^4 + 3z^5) \right. \\
  && \left.\hspace*{144pt} + \, 2 (1 + \lambda_q h) (1-z)^2
            (1 - 2z - 2z^3 + z^4) \right) \\
  && - \, c^4 (1 - z^2) \left( (1 - \lambda_q h)
           (5 - 7z + 7z^2 + 13z^3) \right.\\
  && \left.\hspace*{144pt} - \, 
               2 (1 + \lambda_q h)(1-z)^2 (1 - 4z + z^2) \right) \\
  && + \, c^5 (1 - \lambda_q h) (1-z)^2 (4 + 3z^2 + 4z^3 + 3z^4) \\
  && \left. - \, c^6 (1 - \lambda_q h) (1 + z) (1-z)^3
                         (1 - z + z^2) \right]  \\
 G_A &=& \frac{z(1 - z)}{4 (1 + c + z(1-c))(1 - c + z(1 + c))^3} \\
   &\times& \left[(1 - z^2)(1 + 4z + z^2)
          \left( -(1 - \lambda_q h) z + 2 (1 + \lambda_q h)
                (1 - z) \right) \right.  \\
  && + \, 2c \left( (1 - \lambda_q h) z (1 + 2z + 4z^2 + z^4)
        - 2 (1 + \lambda_q h)(1 + 3z^2 - 3z^3 - z^5) \right)  \\
  && + \, 2c^2 (1 - z) \left( (1 - \lambda_q h) z^2 (1 - 2z - z^2)
       + (1 + \lambda_q h)(1 - z^4) \right) \\
  && - \, 2c^3 z(1 - z) \left( (1 - \lambda_q h)
       (1 + z + 3z^2 + z^3) + 2 (1 + \lambda_q h)(1 - z)^2 \right) \\
  && \left. + \, c^4 (1 - \lambda_q h) z (1 - z)^2 (1 + z^2) \right] \ .
\eean

\vspace{6pt}
\begin{center}
APPENDIX B: THE INVARIANT MASS DISTRIBUTIONS
\end{center}

Here we present the invariant mass distribution of lepton pair
in the massive gluon scheme before factorizing the mass singularity.
Comparing these expressions with those in the dimensional
regularization, we can find the relation between two regularization
schemes.

\vspace{6pt}
\noindent
1. The quark and antiquark annihilation process.
\bean
 \frac{d\hat{\sigma}^{T+V+R}}{dQ^{2}}
   &=& \delta_{\lambda_q , - \lambda_{\bar{q}}} \,
       \frac{4 \pi}{3 N_c} \left( \frac{\alpha}{Q^2} \right)^2
       \left| f^{\lambda_q \lambda} \right|^2 \\
   &\times& z\, \left[ \left( 1 - \frac{7}{4} \,
       \frac{\alpha_s}{\pi} \, C_F \right) \delta(1-z) 
                + \frac{\alpha_{s}}{\pi}
         \left( P_{qq}(z) \ln \frac{Q^2}{\kappa^2} \right.\right.\\ 
   & & \quad + \, \left. \left.
       C_F \left\{ 2 (1 + z^2) 
         \left( \frac{\ln{(1-z)}}{1-z}\right)_+
        - 2 (1 + z^2) \frac{\ln z}{1-z} - 2 (1 - z) \right\}
                     \right) \right] \ .
\eean

\noindent
2. The quark gluon Compton process.
\bean
 \frac{d\hat{\sigma}^C}{dQ^{2}}
   &=& \frac{4 \pi}{3 N_c} \left( \frac{\alpha}{Q^2} \right)^2
       \left| f^{\lambda_q \lambda} \right|^2 \, 
     z\, \left( \frac{\alpha_s}{2\pi} \right)\, 
         \left[ P^C_{qg} (z; \lambda_q , h)
    \left( \ln \frac{Q^2}{\kappa^2} + \ln\frac{1-z}{z^2}  \right)
                                 \right. \\
    & & \hspace*{144pt}
       \left. -\, 
         \frac{1}{8} \left( 1 - 2 z + 3 z^2 + \lambda_q h
            \left\{ 3 - 2 z - 3 z^2 \right\} \right)\right] \ .
\eean
%


\end{document}